\documentclass[preprint,12pt]{elsarticle}

\usepackage{amssymb}
\usepackage{amsthm}
\usepackage{graphicx}
\usepackage{natbib}
\usepackage{color}
\usepackage{lineno}
\usepackage{float}
\usepackage{amsthm}
\usepackage[square,numbers]{natbib}
\usepackage[colorlinks=true,citecolor=blue]{hyperref}
\usepackage{subfig}
\usepackage[english]{babel}

\usepackage{amsmath, amsfonts, amssymb, mathrsfs}

\journal{}

\begin{document}
\numberwithin{equation}{section}
\newtheorem{mydef}{Definition}
\newtheorem{thm}{Theorem}
\newtheorem{lemma}{Lemma}
\newtheorem{prop}{Proposition}
\newtheorem{corol}{Corollary}

\begin{frontmatter}

\title{Impact of intervention on the spread of COVID-19 in India: A model based study}

\author[ISI]{Abhishek Senapati}

\author[VBU]{Sourav Rana\footnote{Corresponding author. Email: sourav.rana@visva-bharati.ac.in}}

\author[VBU]{Tamalendu Das}

\author[ISI]{Joydev Chattopadhyay}

\address[ISI]{Agricultural and Ecological Research Unit, Indian Statistical Institute, 203, B. T. Road, Kolkata 700108, India}

\address[VBU]{Department of Statistics, Visva-Bharati University, Santiniketan, West Bengal, India}

\begin{abstract}
The outbreak of corona virus disease $2019$ (COVID-19), caused by the virus severe acute respiratory syndrome coronavirus 2 (SARS-CoV-2) has already created emergency situations in almost every country of the world. The disease spreads all over the world within a very short period of time after its first identification in Wuhan, China in December, 2019. In India, the outbreaks starts on $2^{nd}$ March, 2020 and after that the cases are increasing exponentially. Very high population density, the unavailability of specific medicines or vaccines, insufficient evidences regarding the transmission mechanism of the disease also make it difficult to fight against the disease properly in India. Mathematical models have been used to predict the disease dynamics and also to assess the efficiency of the intervention strategies in reducing the disease burden. In this work, we propose a mathematical model to describe the disease transmission mechanism between the individuals. We consider the initial phase of the outbreak situation in India and our proposed model is fitted to the daily cumulative new reported cases during the period $2^{nd}$ March, 2020 to $24^{th}$ March, 2020. We estimate the basic reproduction number ($\mathcal{R}_{0}$), effective reproduction number ($R(t)$) and epidemic doubling time from the incidence data for the above-mentioned period. We further assess the effect of preventive measures such as spread of awareness, lock-down, proper hand sanitization, etc. in reducing the new cases. Two intervention scenarios are considered depending on the variability of the intervention strength over the period of implementation. Our study suggests that higher intervention effort is required to control the disease outbreak within a shorter period of time in India. Moreover, our analysis reveals that the strength of the intervention should be strengthened over the time to eradicate the disease effectively.

\end{abstract}


\end{frontmatter}

%\linenumbers
%% main text

\section{Introduction}
Coronaviruses, related to the family \textit{Coronaviridae} are single-stranded, positive-sense RNA viruses \citep{Chen2020a}. In the year $1960$, the human coronaviruses were first classified and scientists used the term `corona' due to its crown-like visibility on the surface area \citep{Kahn2005}. Three major outbreaks occurred because of coronaviruses in the years $2003$, $2012$ and $2015$. The outbreak occurred in China in the year $2003$ was named as `Severe Acute Respiratory Syndrome' (SARS) outbreak. Saudi Arabia in $2012$~\citep{Killerby2020} and South Korea in $2015$~\citep{Willman2019} were affected by 'Middle East Respiratory Syndrome' (MERS) outbreak.

Recently, in December, $2019$, a newly invaded coronavirus, first identified in Wuhan, the capital city of the Hubei province of China caused an outbreak~\citep{Cohen2020}. This is considered as the fourth coronavirus outbreak. It has been reported that due to viral pneumonia, $27$ people were infected which includes $7$ critically ill cases \citep{Tang2020}, and this outbreak has grabbed substantial attentions all over the globe. The Chinese officials declared on $7^{th}$ January, $2020$ that this infectious disease is transmitted through a biological pathogen, named novel coronavirus and on $10^{th}$ January, $2020$, the World Health Organization (WHO) referred the novel coronavirus as `2019-nCoV'. Finally, on $11^{th}$ February, $2020$, WHO declared the official name of the disease and the causing virus as coronavirus disease $2019$ (COVID-19) and severe acute respiratory syndrome coronavirus 2 (SARS-CoV-2) respectively~\citep{WHO2020e}. WHO also published preliminary guidelines for all the countries to deal with this outbreak. The guidelines are mainly based on how to provide health care support to the infected patient and build awareness among the ordinary people about this new disease \citep{WHO2020a}.

Plenty of research works indicates that COVID-19 possibly spread from animal to human (zoonotic) \citep{Linton2020,Rothan2020}. Moreover, a rapid number of new occurrences of COVID-19 reveals the key fact that the secondary transmission can occur through human to human \citep{Li2020} via direct communication or through droplets spread by coughing or sneezing from an infected person \citep{Hui2020}. This human to human transmission is considerably growing almost everywhere in the world by means of the international movement \citep{ECDPC2020} and as soon as an area of a specific country is affected then COVID-19 rapidly grows through local transmission. Person infected with SARS-CoV-2 virus shows symptoms like fever, cough, and shortness of breath, muscle ache, confusion, headache, sore throat, rhinorrhea, chest pain, haemoptysis, diarrhoea, dyspnoea and nausea and vomiting \citep{Chen2020b,Hui2020,Rothan2020}. WHO situation report~\citep{WHO2020f} reveals that the COVID-19 pandemic spread on the Western Pacific, European, South-east Asia, Eastern Mediterranean, America and African Region with their respective territories, causing a huge number of infected cases. The new cases are quickly growing in various countries like USA, Spain, Germany, France, Italy, UK, Iran, Switzerland, India, Netherlands, Austria, etc. \citep{WHO2020f}.

In India, the first confirmed case of COVID-19 was reported on $30^{th}$ January, $2020$~\citep{WHO2020c}. Government of India declared a countrywide lock-down for $21$ days as a preventive measure for the COVID-19 outbreak on $24^{th}$ March, $2020$~\citep{Pulla2020}. Besides the lock-down, the Ministry of Health and Family Welfare (MOHFW) of India, suggested various individual hygiene measures e.g. frequent hand washing, social distancing, use of mask, avoiding touching eyes, nose, or mouth, etc.~\citep{MOHFW2020,Khanna2020}. The government also continuously using various media and social networking web sites to aware the citizen. However, the factors like very high population density, the unavailability of specific medicines or vaccines, insufficient evidences regarding the transmission mechanism of the disease also make it difficult to fight against the disease properly in India.

Mathematical models have been used to predict the disease dynamics and also to assess the efficiency of the intervention strategies in reducing the disease burden. In this context, several studies have been done using real incidence data of the affected countries and examined different characteristics of the outbreak as well as evaluated the effect of intervention strategies implemented to curb the outbreak in the respective countries~\citep{Tang2020,Kucharski2020,Wang2020,Ferretti2020, Zhao2020,Zhao2020a}.

In this work, we propose a deterministic compartmental model to describe the disease transmission mechanism between the individuals. We consider the situation of India during the initial outbreak period and fitted our model to the daily cumulative new cases reported between $2^{nd}$ March, $2020$ to $24^{th}$ March, $2020$. We estimate the basic reproduction number, effective reproduction number and epidemic doubling time from the incidence data for the above-mentioned period. The efficiency of preventive measures in reducing the disease burden is also studied for different level of intervention strength. Further, the impact of intervention in the situation where the strength of the intervention is varied over the implementation period.

The rest of the paper is organized as follows. In section~\ref{sec:model_formulation}, we briefly describe our proposed model. Section~\ref{sec:model_fitting} is devoted in describing the procedure of model fitting. The estimation of basic reproduction number, effective reproduction number and epidemic doubling time from actual incidence data is described in section~\ref{sec:basic_reproduction}. In section~\ref{sec:intervention}, the efficiency of the intervention is studied. Finally we discuss the findings obtained from our study in section~\ref{sec:discussion}.

\section{Description of the model} \label{sec:model_formulation}

We adopt deterministic compartmental modelling approach to describe the disease transmission mechanism. Depending on the health status, the total human population is categorized into seven compartments: susceptible ($S$), exposed ($E$), symptomatic ($I$), asymptomatic ($I_{a}$), quarantined ($I_q$), hospitalized ($H$) and recovered ($R$). Susceptible population becomes exposed with the disease after experiencing close contacts with the symptomatic as well as asymptomatic individuals. We assume that the rate of disease transmission from asymptomatic individuals to susceptible individuals is less than that of from symptomatic individuals. The rate of new infection is given by $\frac{\beta S(I+\eta I_{a})}{N}$, where $\beta$ denotes the transmission rate of the disease and $\eta(<1)$ is the modification parameter that accounts the reduction in the transmission rate from the asymptomatic individuals. At any instant of time, the total population is given by $N=S+E+I+I_{a}+I_{q}+H+R$. Since we consider the outbreak situation which usually persists for a shorter period of time, we do not incorporate any demographic factors (i.e birth, death, etc.) into the model. We assume that after the incubation period ($\sigma^{-1}$), $\rho_{1}$ fraction of the exposed individuals move to symptomatic compartment, $\rho_{2}$ fraction move to the asymptomatic compartment and the remaining fraction, $(\rho_{3}=1-\rho_{1}-\rho_{2})$ move to the quarantined compartment. The individuals in the symptomatic compartment ($I$) show severe symptoms of the disease and after $\alpha^{-1}$ period of time they are hospitalized. On the other hand, the asymptomatic individuals who do not show any symptom of the disease get natural recovery at a rate $\gamma_{a}$ and move to the recovered class $(R)$. The individuals in the quarantined compartment ($I_{q}$) are those individuals who exhibit mild symptoms and are advised to be quarantined. From quarantined class ($I_{q}$), individuals move to the hospitalized class ($H$) at a rate $\alpha_{q}$. They can also get natural recovery and move to the recovered class ($R$) at a rate $\gamma_{q}$. Individuals admitted in the hospitals move to recovered class at a rate $\gamma$. We also consider that the hospitalized individuals die due to the disease at a rate $\delta$. The recovered population increases due to the recovery of asymptomatic, quarantined and hospitalized individuals at the rates $\gamma_{a}$, $\gamma_{q}$ and $\gamma$ respectively. The following set of ordinary differential equations represents the transmission dynamics of the disease:

\begin{eqnarray}
%\small
\begin{array}{llll}
\displaystyle \frac{d S}{dt} &=& \displaystyle - \frac{\beta S(I+\eta I_{a})}{N}, \\\\
\displaystyle \frac{d E}{dt} &=& \displaystyle  \frac{\beta S(I+\eta I_{a})}{N} -\sigma E,\\\\
\displaystyle \frac{d I}{dt} &=& \displaystyle  \rho_{1}\sigma E -\alpha I ,\\\\
\displaystyle \frac{d I_{a}}{dt} &=& \displaystyle \rho_{2}\sigma E -\gamma_{a} I_{a} ,\\\\
\displaystyle \frac{d I_{q}}{dt} &=& \displaystyle  \rho_{3}\sigma E -(\alpha_{q}+\gamma_{q})I_{q},\\\\
\displaystyle \frac{d H}{dt} &=& \displaystyle  \alpha I + \alpha_{q}I_{q}-(\gamma +\delta) H,\\\\
\displaystyle \frac{d R}{dt} &=&  \gamma_{a} I_{a} + \gamma_{q} I_{q}+\gamma H.
\end{array}
\label{EQ:Model}
\end{eqnarray}

\begin{figure}[H]
	\begin{center}
		{\includegraphics[width=1\textwidth]{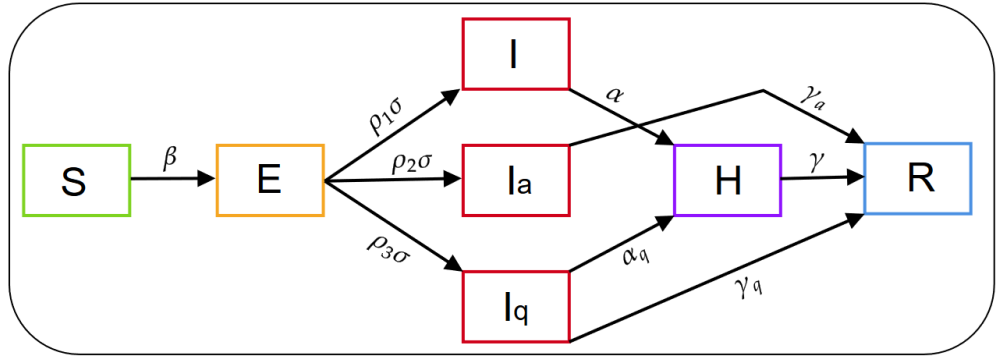}}
	\end{center}
	\caption{Schematic diagram of the model~\eqref{EQ:Model}.}
	\label{fig:schematic}
\end{figure}

The schematic diagram and the description of the parameters used in the model~\eqref{EQ:Model} is presented in Fig.~\ref{fig:schematic} and Table~\ref{table-1} respectively.

%%%%%%%%%%%%%%%%%%%% PARAMETERS
\begin{table}[h!]
\tabcolsep 7pt
\centering
\begin{tabular}{c p{5cm} p{4cm} p{2cm}}
\hline
\tiny{\textbf{Parameters}} & \tiny{\textbf{Description}} &\tiny{\textbf{Value}} & \tiny{\textbf{Reference}}  \\
[0.5ex]
\hline
\tiny{$\beta$} & \tiny{Rate of disease transmission} & \tiny{$1.1164$} & \tiny{Estimated}\\

\tiny{$\eta$} & \tiny{Modification parameter} & \tiny{$0.2$} & \tiny{Estimated}\\

\tiny{$\sigma$} & \tiny{Rate of transition from exposed to infected class} & \tiny{$0.1923~day^{-1}$} & \tiny{\citep{Li2020}}\\

\tiny{$\rho_{1},\rho_{2},\rho_{3}$} & \tiny{Fractions of population move from the compartment $E$ to the compartments $I$, $I_{a}$ and $I_{q}$ respectively} & \tiny{$0.5210$, $0.2740$, $0.2050$} & \tiny{Estimated}\\

\tiny{$\alpha$} & \tiny{Rate of transition from symptomatic to hospitalized class} & \tiny{$0.2174~day^{-1}$} & \tiny{\citep{Li2020}}\\

\tiny{$\alpha_{q}$} & \tiny{Rate of transition from quarantined to hospitalized class} & \tiny{$ 0.1429~day^{-1}$} & \tiny{Assumed}\\

\tiny{$\gamma$} & \tiny{Recovery rate of the individuals in hospitalized compartment} & \tiny{$ 0.0701~day^{-1}$} & \tiny{\citep{Covid19}}  \\

\tiny{$\gamma_{a}$} & \tiny{Recovery rate of asymptomatic population} & \tiny{$0.13978~day^{-1}$} & \tiny{\citep{Tang2020}}  \\

\tiny{$\gamma_{q}$} & \tiny{Recovery rate of quarantined population} & \tiny{$0.11624~day^{-1}$} & \tiny{\citep{Tang2020}}  \\

\tiny{$\delta$} & \tiny{Rate of disease induced death} & \tiny{$0.0175~day^{-1}$} & \tiny{\citep{Covid19}}  \\
[1ex]
\hline
\end{tabular}
\caption{\small{\textbf{Description of parameters used in the model~(\ref{EQ:Model}).}}}
\label{table-1}
\end{table}

\section{Model fitting}
\label{sec:model_fitting}

Though the first confirmed case of COVID-19 was reported on $30^{th}$ January, $2020$~\citep{WHO2020c}, from $2^{nd}$ March, 2020 onwards, the new cases are being reported continuously. Therefore, we consider $2^{nd}$ March, $2020$ as the starting date of the outbreak in India. It is also to be noted that, Government of India declared a countrywide lock-down from $25^{th}$ March, $2020$ for $21$ days~\citep{Pulla2020}. Therefore, we assume that during the period $2^{nd}$ March, $2020$ to $24^{th}$ March, $2020$ no such preventive measures was taken by the Government of India. Since we do not incorporate any intervention in our model~\eqref{EQ:Model}, we fit our model to the daily cumulative new reported COVID-19 cases of India during the period $2^{nd}$ March, $2020$ to $24^{th}$ March, $2020$. The daily cumulative cases data are obtained from~\citep{JohnsHopkins20}.

We estimate four unknown model parameters: $(i.)$ the transmission rate ($\beta$), $(ii.)$ modification parameter ($\eta$), $(iii.)$ fraction of population move to symptomatic class from exposed class ($\rho_{1}$) and $(iv.)$ fraction of population move to asymptomatic class from exposed class ($\rho_{2}$) by fitting the model to the cumulative reported cases data.

The cumulative new reported cases from the model is given by

\begin{equation}
C(t,\Theta)=C(0)+\int_{0}^{t} (\alpha I(\tau)+\alpha_{q}I_{q}(\tau)) d\tau,
\end{equation}
where $\Theta=\{\beta, \eta, \rho_{1}, \rho_{2}\}$ and $C(0)$ denotes the initial cumulative cases.

We perform our model fitting by using in-built function \textit{lsqnonlin} in MATLAB (Mathworks, R2014a) to minimize the sum of square function. In our case, the sum of square function $SS(\Theta)$ is given by,

\begin{equation}
SS(\Theta)=\displaystyle \sum_{i =i}^{n} \big(C^{d}(t_{i})-C(t_{i},\Theta)\big)^2, \\\nonumber
\end{equation}
where, $C^{d}(t_{i})$ is the actual data at $t_{i}^{th}$ day and $n$ is the number of data points.

\begin{figure}[H]
	\begin{center}
		{\includegraphics[width=1\textwidth]{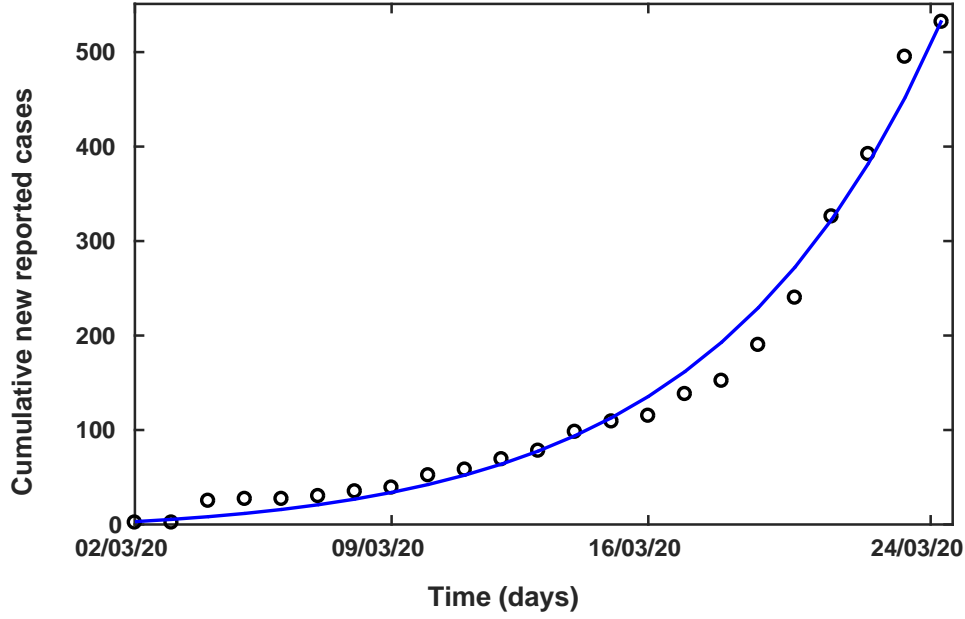}}
	\end{center}		
	
	\caption{\textbf{Model fitting to cumulative new COVID-19 reported cases for the time period $2^{nd}$ March, 2020 to $24^{th}$ March, 2020. The blue solid line represents the model solution and black circles are the discrete data points. }}
	\label{fig:fitting_cum_new_cases}
	
\end{figure}

The model fitting to the cumulative new reported cases is displayed in Fig.~\ref{fig:fitting_cum_new_cases}. The values of the estimated parameters are given in Table~\ref{table-1}.

\section{Basic reproduction number}
\label{sec:basic_reproduction}

Basic reproduction number is a key quantity in epidemiology which quantifies the average number of secondary cases generated from a single primary cases during the his infectious period. This quantity can be calculated from a mathematical model by following next generation matrix approach. Following~\citep{van2002reproduction}, the new infection matrix $F$ and the transmission matrix $V$ are given by,

\[ F = \left( \begin{array}{ccccccc}
0  &  \beta & \eta \beta & 0 \\
0  & 0  & 0 & 0 \\
0  & 0  & 0 & 0\\
0  & 0  & 0 & 0 \end{array} \right), V = \left( \begin{array}{ccccccc}
\sigma  & 0 & 0 & 0 \\
-\rho_{1}\sigma  &  \alpha & 0 & 0 \\
-\rho_{2}\sigma  &  0  & \gamma_{a} & 0   \\
-\rho_{3}\sigma  &  0  & 0   & \alpha_{q}+\gamma_{q}  \end{array} \right).\]

The basic reproduction number, $\mathcal{R}_{0}$ is defined as the spectral radius of the matrix $FV^{-1}$,

\begin{eqnarray}
\centering
\begin{array}{llll}
\mathcal{R}_0 &=& \displaystyle{\frac{\rho_1\beta}{\alpha}+\frac{\eta\rho_2\beta}{\gamma_a}}.\\
\end{array}
\label{Eq:Basic_Reproduction}
\end{eqnarray}

It is to be noted that, the first term is the number of new infection caused by symptomatic individual ($I$) whereas the second term is the number new infection due to the infection by asymptomatic individual ($I_{a}$).

\subsection{Estimation of basic reproduction number from actual epidemics}
The basic reproduction number $(\mathcal{R}_{0})$ for communicable diseases can be estimated through the actual epidemic data by using various statistical as well as mathematical methods \citep{Massad10}. In this study, we estimate $\mathcal{R}_{0}$ from the initial growth phase of the epidemics \citep{Massad01,Sardar2015,Pinho10}. At the early stage of the epidemic, there is a non-linear relationship between the cumulative number of cases $C(t)$ and the force of infection $\Lambda$ which can be mathematically written as $C(t) \propto \exp(\Lambda t)$. So the number of exposed, symptomatic and asymptomatic population progress in the following form,
\begin{eqnarray}
\centering
\begin{array}{llll}
E & \propto & E_0 \exp(\Lambda t),\\
I & \propto & I_0 \exp(\Lambda t),\\
I_a & \propto & I_{a0} \exp(\Lambda t).
\end{array}
\label{Eq:Force_of_Infection_1}
\end{eqnarray}
where $E_0$, $I_0$ and $I_{a0}$ are constants. Further, the number of non-susceptible population can be assumed negligible i.e. $S(t)= N$. Substituting the equation \eqref{Eq:Force_of_Infection_1} into the model \eqref{EQ:Model}, we have
\begin{eqnarray}
\centering
\begin{array}{llll}
E_0(\Lambda + \sigma) & = & \beta (I_0+\eta I_{a0}),\\
I_0(\Lambda + \alpha) & = & \rho_1 \sigma E_0,\\
I_{a0}(\Lambda + \gamma_a) & = & \rho_2 \sigma E_0.
\end{array}
\label{Eq:Force_of_Infection_2}
\end{eqnarray}

Using the expression of $\mathcal{R}_{0}$ from equation \eqref{Eq:Basic_Reproduction} and applying the above equations \eqref{Eq:Force_of_Infection_2} we have determined relation between the $\mathcal{R}_{0}$ and the force of infection $\Lambda$ as follows:

\begin{eqnarray}
\centering
\begin{array}{llll}
\mathcal{R}_{0} & = & \displaystyle{\Bigr(\frac{\rho_1}{\alpha}+\frac{\rho_2 \eta}{\gamma_a}\Bigr) \Biggr[ \frac{\Lambda + \sigma}{\frac{\rho_1 \sigma}{\Lambda+ \alpha}+\frac{\eta \rho_2 \sigma}{\Lambda+\gamma_a}}  \Biggr]}.
\end{array}
\label{Eq:Force_of_Infection_3}
\end{eqnarray}

Here we first estimate the force of infection $\Lambda$ and then estimate $\mathcal{R}_{0}$ by using equation \eqref{Eq:Force_of_Infection_3}. Following \cite{Favier06,Pinho10}, the relation between the number of new cases per day and the cumulative number of cases per day $C(t)$ as: number of new cases $\sim$ $\Lambda C(t)$.

The force of infection ($\Lambda$) can be calculated from the COVID-19 incidence data in the following ways \citep{Sardar2015}:

%\begin{description}[nolistsep]
\begin{itemize}

	\item[Step~1.] We plot the number of new COVID-19 cases (per day) in $x-axis$ versus the cumulative number of COVID-19 cases (per day) in $y-axis$.
	\item[Step~2.] In the scatter plot, we point out the threshold of cumulative cases up to which new cases show the exponential growth.
	\item[Step~3.] Then we fit a linear regression model using the least square technique to this exponential growth data.
	\item[Step~4.] The slope of the fitted line is considered as the force of infection ($\Lambda$).
\end{itemize}
%\end{description}

\begin{figure}[H]
	\begin{center}
		{\includegraphics[width=1\textwidth]{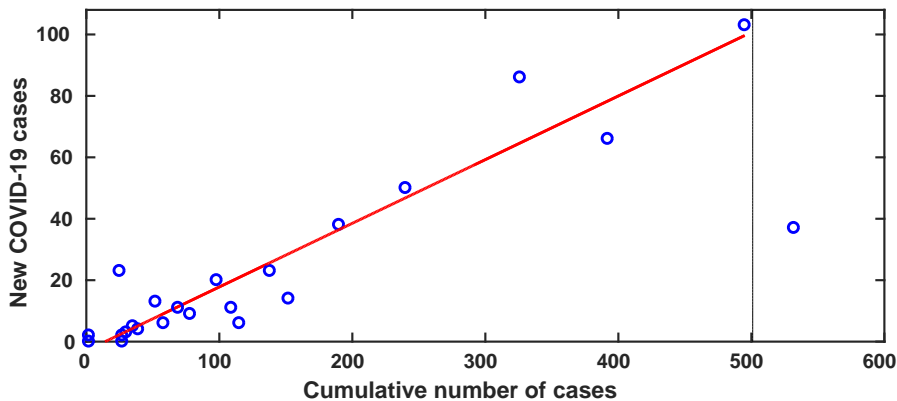}}
	\end{center}		
	\caption{\textbf{The daily number of new COVID-19 cases against the cumulative number of COVID-19 cases in India,
    from $2^{nd}$ March, 2020 to $24^{th}$ March, 2020. The box indicates the growing linear parts of the plots corresponding to the initial growth of the epidemics. The least squares linear fit of the linear phase corresponding to data in dot box gives $\Lambda = 0.2073 \pm 0.0151$ with correlation coefficient $R \approx 0.9507$.}}
	\label{fig:force_of_infection}	
\end{figure}

We obtain $\Lambda = 0.2073 \pm 0.0151~day^{-1}$ based on the slope shown in Fig.~\ref{fig:force_of_infection}. Using the equation \eqref{Eq:Force_of_Infection_3} along with the parameter values form Table \ref{table-1}, we obtain the estimate of $\mathcal{R}_{0} = 4.1849$ with upper and lower bounds $4.5014$ and $3.8799$ respectively. This estimate of $\mathcal{R}_0$ indicates that the initial transmissibility of COVID-19 is pretty much higher than $1$, which in turns implies that it is essential to control the disease at the initial phase.

\subsubsection{Epidemic doubling time}

The doubling time of an epidemic is a measure of the rate of spread of a disease. It is the required time to double the number of cases in the epidemic. There is an inverse relationship between epidemic strength and the doubling time i.e. if an epidemic declines, the doubling time increases and vice versa. Following \citep{Ferretti2020}, we obtain the epidemic doubling time for our study is $T_2=\frac{\ln(2)}{\Lambda}\approx 3.34$ days. Therefore, it is necessary to apply some preventive measures else the epidemic appears in a large scale within a short time.

\subsection{Effective reproduction number~$R(t)$}\label{Effective-reproduction-Rt}
Study regarding the time span of an epidemic is very crucial and can be achieved to a certain extent through the estimation of $R(t)$, the effective reproduction number. It is defined as the actual average number of secondary cases from a typical primary case at time $t$ \citep{Nishiura2009}. The value of $R(t)$ provides information about the severity of the disease over different time points and alert epidemiologists to suggest about the control measures \citep{Sardar2015}. We estimate $R(t)$ from the daily new infection curve of the infected COVID-19 cases data by using the following equation derived from the renewal equation of a birth process:
\begin{eqnarray}
\centering
\begin{array}{llll}
R(t) &=& \displaystyle\frac{b(t)}{\int_{a = 0}^{\infty} b(t-a) g(a)
	\textup{d}a},\\
\end{array}
\label{Renewal-equation}
\end{eqnarray}where, the term $b(t)$ corresponds to the number of new cases in the day $t$ and the term $g()$ is the generation interval distribution for a disease \citep{Pinho10}.

We derive the expression of the generation interval distribution $g(t)$ from the model~\eqref{EQ:Model} by applying the method discussed in \cite{Pinho10,Wallinga07}. The rates of leaving the exposed and infectious compartments are indicated by $b_{1},b_{2}$ and $b_{3}$. These quantities are constant and extracted from the model~\eqref{EQ:Model} as $b_{1} = \sigma$, $b_{2} = \alpha$ and $b_{3} = \gamma_a$. Moreover, the generation interval distribution is the convolution of three exponential distributions with a mean $T_{c} = 1/b_{1}+1/b_{2}+1/b_{3}$. Following~\cite{Pinho10,Akkouchi08}, we have the following explicit expression for the convolution:

\begin{eqnarray}
\centering
\begin{array}{llll}
g(t) &=& \displaystyle\sum_{i = 1}^{3} \frac{b_{1} b_{2} b_{3}  \exp{(- b_{i}t)}}{\displaystyle\prod_{j =1, j \neq i}^{3} (b_{j} - b_{i})},\\
\end{array}
\label{generation-interval-distribution}
\end{eqnarray}with $t \geq 0$. The validity of the above relation~\eqref{generation-interval-distribution} holds for a minimum threshold value of the force of infection $\Lambda$, defined as $\Lambda > \min{(- b_{1}, - b_{2},- b_{3})}$ \citep{Wallinga07}.

\begin{figure}[H]
	\begin{center}
		{\includegraphics[width=1 \textwidth]{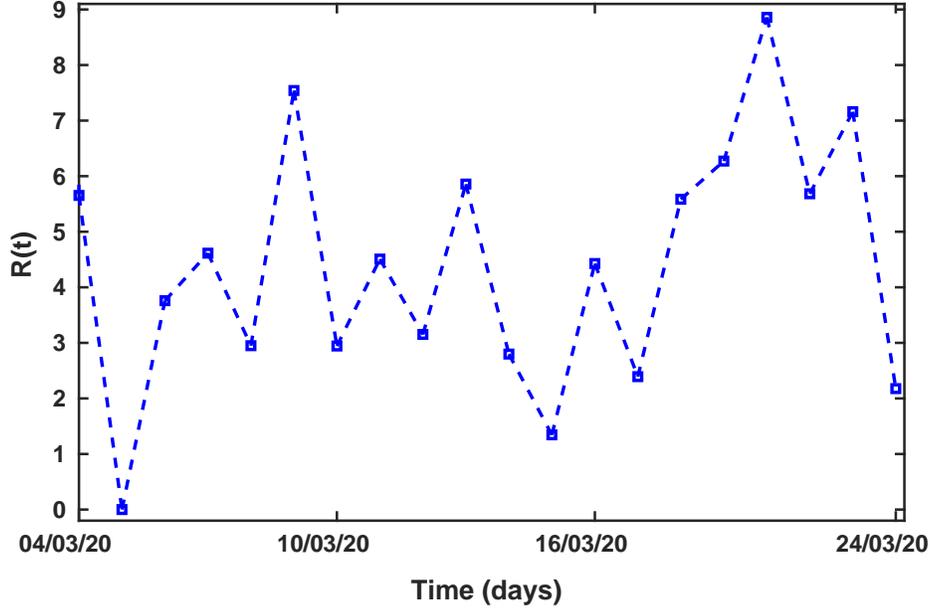}}
	\end{center}	
	\caption{\textbf{The effective reproductive number $R(t)$ versus time $t$ (days) of the COVID-19 outbreak in India during $2^{nd}$ March, $2020$ to $24^{th}$ March, $2020$. The parameter values are taken from Table \ref{table-1}.}}
	\label{fig:effective_R0}	
\end{figure}

Using the daily COVID-19 incidence data and applying the expression of $g(t)$ in equation \eqref{generation-interval-distribution}, we estimate $R(t)$ from equation~(\ref{Renewal-equation}). Fig. \ref{fig:effective_R0} shows the time evolution of the effective reproductive number $R(t)$ to the COVID-19 outbreak in India, from $2^{nd}$ March, $2020$ to $24^{th}$ March, $2020$. The result is shown here for $t \geq 3$, since, the method used here to derive the expression of $R(t)$ is not applicable for lower values of $t$. Here the value of $R(t)$ lies between $2$ to $6$ most of the times. The low value of $R(t)=0$ appears on the second point (see Fig. \ref{fig:effective_R0}) due to non occurrence of new cases at $6^{th}$ March $2020$ also the high value of $R(t)\approx 7.54,~8.86$ occurs on $6^{th},17^{th}$ day of the outbreak due to the high number of new cases found on these days. It implies that the disease continues to infect the more and more people during that period. It is worthy to remember that, at this point Govt. of India announced a lock-down to break the chain of infection spread else it will be almost impossible to control the spread after a certain period of time.

\section{Effect of intervention}
\label{sec:intervention}

In this section, we study the effect of intervention strategies in reducing the new COVID-19 cases through our model. Intervention strategies includes the control measures such as lock-down, spreading awareness program through media, proper hand sanitization, etc. which results in slowing down the disease transmission process. In terms of model parameter, the implementation of intervention implies that there would be a reduction in the disease transmission rate $\beta$. This reduction in the transmission rate is considered as the strength of the intervention, $k$. In the presence of intervention, the parameter $\beta$ is modified as $(1-k)\beta$ throughout the period of implementation.

Staring from the initial date of outbreak, we consider a period of $210$ days (i.e $7$ months) and study the impact of intervention during that period. Speaking in terms of actual date, we consider the time period $2^{nd}$, March, $2020$ to $27^{th}$ September, $2020$.

We consider two types of intervention scenarios: $(i.)$ the strength of the intervention ($k$) is fixed throughout the implementation period, $(ii.)$  the strength of intervention is varied over the implementation period.

\subsection{Scenario 1}
\begin{figure}[H]
	\begin{center}
		{\includegraphics[width=1\textwidth]{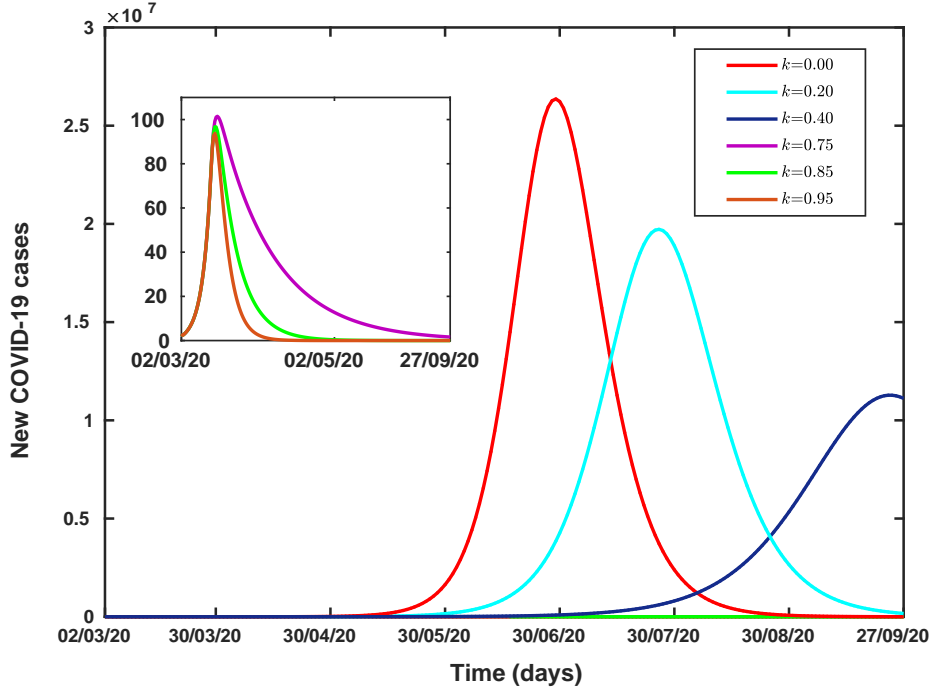}}
	\end{center}		
	
	\caption{\textbf{Trajectories of the new COVID-19 cases for different values of the strength of the intervention ($k$). The zoomed in figures for $k=0.75$, $0.85$ and $0.95$ are displayed in the insets. All the parameters are taken from Table~\ref{table-1}. }}
	\label{fig:control_newcases}
	
\end{figure}

In this intervention scenario, the strength of the intervention $k$ is taken to be fixed over the implementation period. We first consider the situation without any intervention (i.e $k=0$). Our model~\eqref{EQ:Model} is simulated for $210$ days to observe the dynamics of the disease in the absence of any intervention, using the in-built function \textit{ode45} in MATLAB (Mathworks, R2014a). From Fig.~\ref{fig:control_newcases}, we see that the number new cases grows exponentially and attains the maximum at the end of the June. It is also observed that maximum of $26.3593$ million new cases can occur in the absence of any intervention during the period (see Table~\ref{table-2}). However, in the subsequent time, decreasing trend in the new cases is noted.

Now we study the efficiency of the intervention by varying the strength of the intervention ($k$). Regarding the initiation of the control, we follow the same date when Govt. of India implemented nation-wide lock down i.e $25^{th}$ March, 2020. Throughout the study we consider this date as the initial date of implementing the intervention.

For the low value of the intervention strength (i.e for $k=0.20$), it is observed that the time for the occurrence of peak of the outbreak is slightly delayed than that of without intervention scenario (see Fig.~\ref{fig:control_newcases}). The peak of the outbreak is shifted to the end of the month July. In this case, maximum $19.7233$  million new cases can occur (see Table~\ref{table-2}). If the strength of the intervention is increased further (for $k=0.40$), the peak of the outbreak decreases and the occurrence of the peak shifted to the end of September (see Fig.~\ref{fig:control_newcases}). In this case, the maximum number new cases in a single day reduces to $11.2784$ million (see Table~\ref{table-2}).

Now we consider the higher values of intervention strength. For $k=0.75$, the new cases tend to decrease within a week (see Fig.~\ref{fig:control_newcases}) from the date of implementation and in such a case maximum of $102$ new cases can occur (see Table~\ref{table-2}). If the strength is increased further (i.e for $k=0.85$ and $0.95$) we see that the disease can be effectively eradicated within $1$ to $2$ months from the initial date of implementing intervention (see Fig.~\ref{fig:control_newcases}).

\begin{figure}[H]
	\begin{center}
		{\includegraphics[width=1\textwidth]{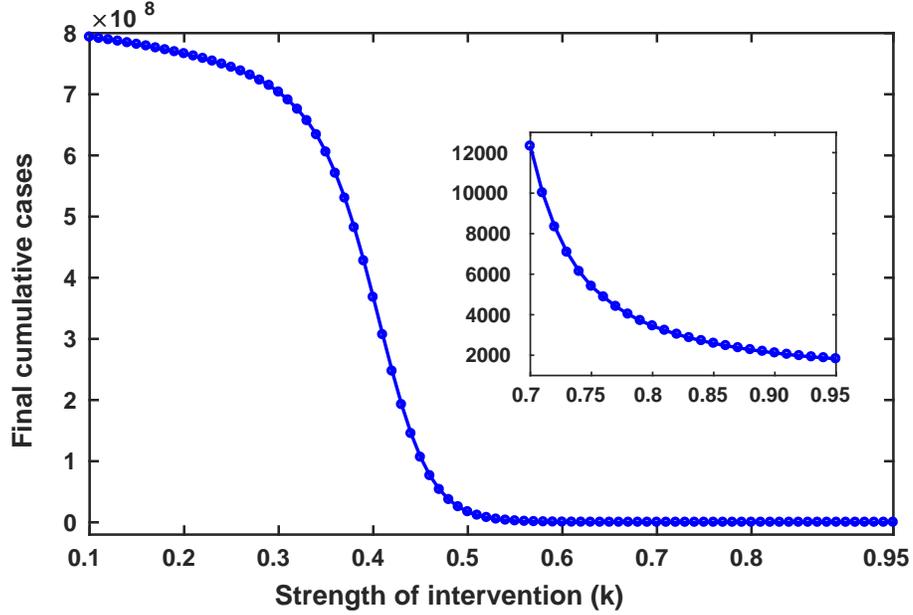}}
	\end{center}		
	
	\caption{\textbf{Final cumulative new cases for different values of strength of intervention ($k$). The zoomed in figure is displayed in the inset. }}
	\label{fig:control_final_cumulative}
	
\end{figure}

Next we quantify the cumulative number of new cases at the end of the time duration under consideration (i.e at the end of $27^{th}$ September, $2020$) for different values of strength of intervention ($k$). From Fig.~\ref{fig:control_final_cumulative}, we see that, the final number of cumulative new cases gradually decreases if the strength of the intervention is increased. However, for higher values of $k$ (i.e for $k\in[0.7,0.95]$), the final cumulative cases are reduced significantly. The actual number of final cumulative cases for different values of $k$ are presented in Table~\ref{table-2}.

\begin{table}[h!]
	\tabcolsep 7pt
	\centering
	\begin{tabular}{ccc}
		\hline
		\tiny{\textbf{Strength of the intervention ($k$)}} & \tiny{\textbf{Final cumulative cases}} &\tiny{\textbf{Maximum number of new cases}}   \\
		[0.5ex]
		\hline
			\tiny{$0.0$} & \tiny{$810.5521$ million} & \tiny{$26.3593$ million}\\
			
		\tiny{$0.2$} & \tiny{$766.2467$ million} & \tiny{$19.7233$ million}\\
		
		\tiny{$0.4$} & \tiny{$367.8458$ million} & \tiny{$11.2784$ million}\\
		
		\tiny{$0.6$} & \tiny{$0.3141$ million} & \tiny{$0.0072$ million}\\

        \tiny{$0.7$} & \tiny{$12359$} & \tiny{$107$}\\
		
		\tiny{$0.75$} & \tiny{$5424$} & \tiny{$102$}\\
		
		\tiny{$0.85$} & \tiny{$2587$} & \tiny{$97$ }\\
		
		\tiny{$0.95$} & \tiny{$1814$} & \tiny{$94$}\\
		[1ex]
		\hline
	\end{tabular}
	\caption{\small{\textbf{Actual number of final cumulative cases and maximum number of new cases for different values of $k$.}}}
	\label{table-2}
\end{table}

\begin{figure}[H]
	\begin{center}
		{\includegraphics[width=1\textwidth]{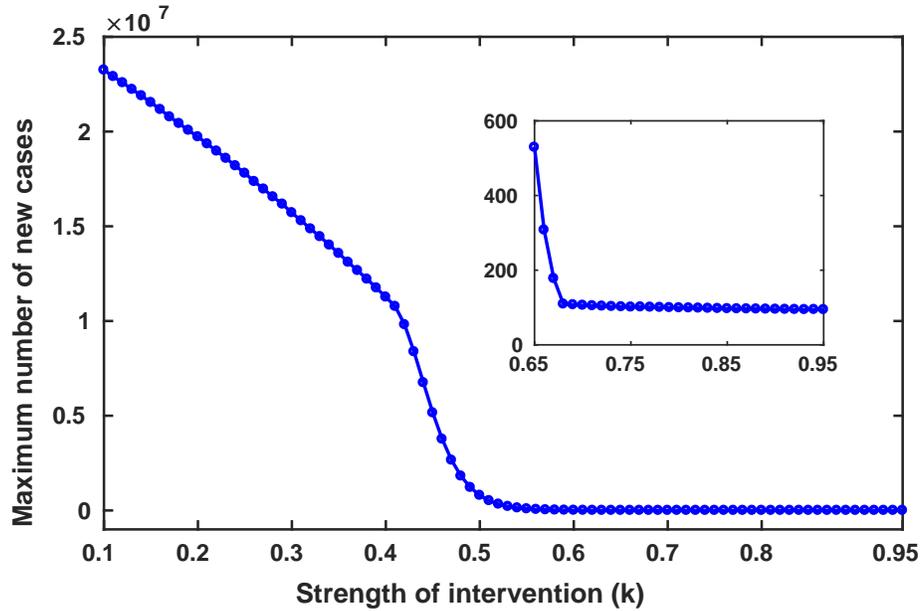}}
	\end{center}		
	
	\caption{\textbf{Maximum number of new cases for different values of strength of intervention ($k$) for the time period $2^{nd}$ March, $2020$ to $27^{th}$ September, $2020$. The zoomed in figure is displayed in the inset.}}
	\label{fig:control_max_new_cases}
	
\end{figure}

Next we evaluate the maximum number of new cases during the time period $2^{nd}$ March, $2020$ to $27^{th}$ September, $2020$ for different values of $k$ (see Fig.~\ref{fig:control_max_new_cases}). The actual number of maximum new cases for different values of $k$ are tabulated in Table~\ref{table-2}.

\begin{figure}[H]
	\begin{center}
		{\includegraphics[width=1\textwidth]{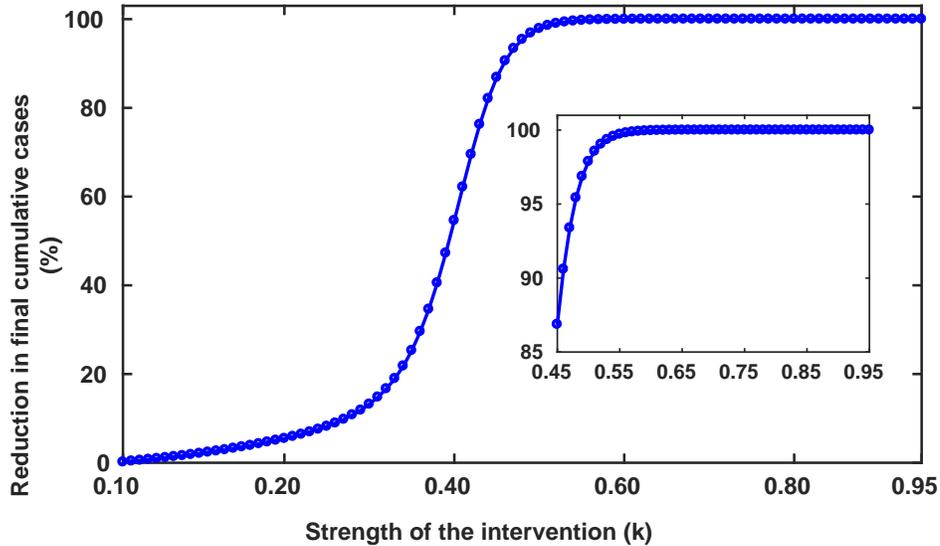}}
	\end{center}		
	
	\caption{\textbf{Percentage of relative reduction in the final cumulative new cases for different values of strength of intervention ($k$).}}
	\label{fig:reduction}
	
\end{figure}

The percentage of relative reduction in the final cumulative new cases for different $k$ is presented in Fig.~\ref{fig:reduction}. It is observed that interventions having lower strengths(i.e for $k\in[0.1,0.4]$), can reduce up to $54.6\%$ final cumulative cases. However, if the strength of the intervention is considered to be higher then almost $99.8\%$ reduction in the final cumulative cases can be achieved.

\subsection{Scenario 2}

In this scenario, instead of considering the fixed value of $k$ over the whole time interval after the initiation of intervention as in scenario 1, we vary the value of $k$ over the implementation period. The implementation period i.e $25^{th}$ March, $2020$ to $27^{th}$ September, $2020$ is divided into three time windows. The first window consists of $21$ days whereas the second and third window consists of $35$ days and $131$ days respectively.

\begin{figure}[H]
	\begin{center}
		{\includegraphics[width=1\textwidth]{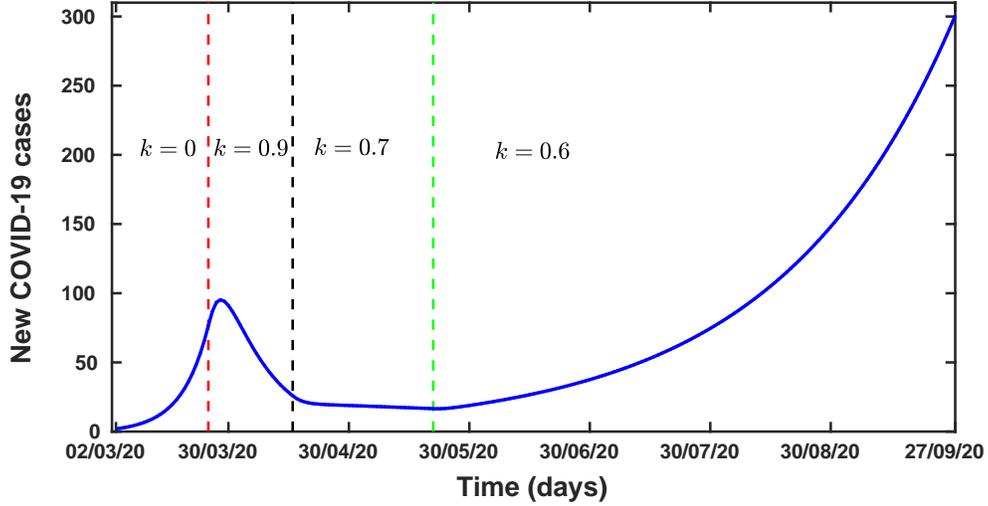}}
	\end{center}		
	
	\caption{\textbf{Time evolution of the new COVID-19 cases for the period $2^{nd}$ March $2020$ to $27^{th}$ September, $2020$. In each time window the strength of the intervention $(k)$ is given. The strength of the intervention is decreased in the subsequent time windows after the initiation of the intervention}}
	\label{fig:high_to_low}

\end{figure}

We first decrease the value of $k$ from first window to third window. The values of $k$ taken in the three windows are $0.9$, $0.7$ and $0.6$ respectively. We observe from Fig.~\ref{fig:high_to_low} that in this case, the new cases decreases in the first window and remains almost unchanged in the second window where the value of $k$ is decreased slightly than first window. However, when the value of $k$ is decreased further in the third window, the new cases tend to increase rapidly. This implies that the relaxation in the interventions strategy over the time does not end up with the disease eradication.

\begin{figure}[H]
	\begin{center}
		{\includegraphics[width=1\textwidth]{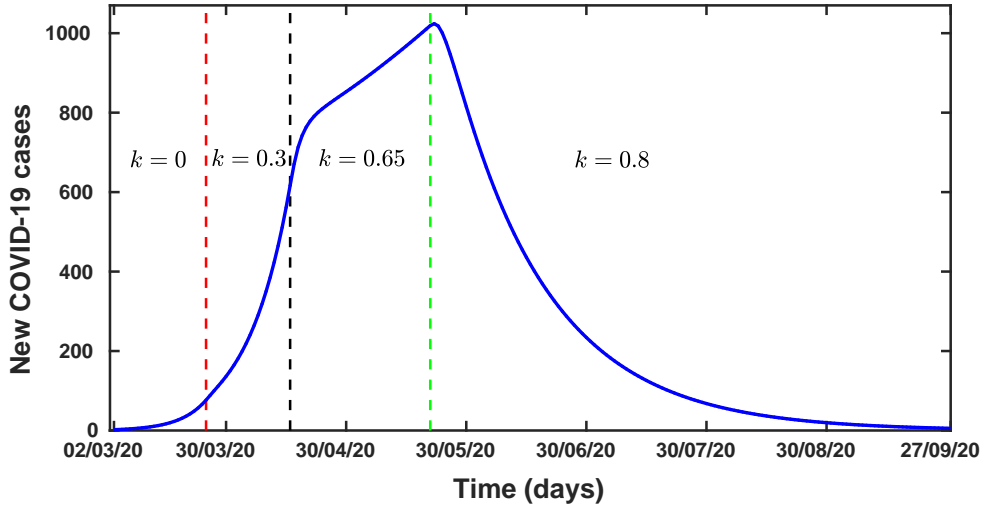}}
	\end{center}		
	
	\caption{\textbf{Time evolution of the new COVID-19 cases for the time period $2^{nd}$ March $2020$ to $27^{th}$ September, $2020$. In each time window the strength of the intervention $(k)$ is given. The strength of the intervention is increased in the subsequent time windows after the initiation of the intervention.}}
	\label{fig:low_to_high}
	
\end{figure}

Next we consider the opposite case i.e we increase the value of $k$ from first window to third window. The values of $k$ are taken as $0.3$, $0.65$ and $0.8$ respectively in the respective time windows. It is observed from Fig.~\ref{fig:low_to_high} that the new cases tends to increase in the first two windows. However, in the third window where the value of $k$ is high enough, the new tend to decrease. This essentially implies that the intervention should be strengthened over the time to eradicate the disease effectively.

\section{Discussion and conclusion}
\label{sec:discussion}

The COVID-19 outbreak in India is a potential threat to the country due to its rapid spread. Mathematical models are very effective tools to predict the time span and pattern of the outbreak. Moreover, mathematical models can also provide useful insights regarding the impact of intervention in lowering the disease incidence.

In this study, we proposed a deterministic compartmental model to describe the disease transmission mechanism among the population. We considered the initial phase of outbreak of the disease COVID-19 in India and fitted our proposed model to the cumulative new reported cases during the period $2^{nd}$ March, $2020$ to $24^{th}$ March, $2020$. Some model parameters are estimated by fitting our model to the cumulative new reported cases during the above mentioned period. By looking at the estimated parameters, it is observed that the rate of disease transmission is quite high which basically implies the high infectiousness of the disease. The percentage of the symptomatic individuals coming from exposed individuals is estimated to be more than $52\%$, whereas the percentage of the asymptomatic and quarantined individuals are estimated as more than $27\%$ and $20\%$ respectively. This indicates that the contribution of the asymptomatic population to COVID-19 cases in India is not negligible.

Based on the estimated parameters and actual COVID-19 incidence data, we estimated basic reproduction number, effective reproduction number and epidemic doubling time to get an overview of this initial phase of outbreak. We obtained the estimate of basic reproduction number $\mathcal{R}_0$ as $4.1849$ with upper and lower bounds are $4.5014$ and $3.8799$ respectively. This high value of $\mathcal{R}_{0}$ basically captures the outbreak scenario in India. The effective reproduction number ($R(t)$) provides information about the severity of COVID-19 over different time points. In our study, the values of $R(t)$ lie between $2$ and $6$ most of the time. This is also confirms high transmissibility of the disease. The epidemic doubling time is also estimated to be approximately $3.34$ days. This suggests that the rate of disease transmission need to be controlled otherwise a large proportion will be affected within a very short period of time

We studied the impact of intervention in reducing the disease burden. We basically considered the preventive measures such as lock-down, spreading of awareness program through media, proper hand sanitization, etc. which slow down the disease transmissibility. Two intervention scenarios are considered depending on the variability of the intervention strength over the period of implementation. In the first scenario, we fixed the strength of the intervention throughout the period of implementation and  studied the impact of intervention for different level of intervention efforts. In this scenario, our study reveals that higher intervention effort is required to control the disease outbreak within a shorter period of time. In the second scenario, the whole implementation time are divided into three time windows and in each of the window, the intervention strength is taken to be different. In such a scenario, our analysis shows that the strength of the intervention should not be relaxed over the time rather the intervention should be strengthened to eradicate the disease effectively. Designing the efficient intervention strategy is one of the crucial factor to curb the disease spread in an outbreak situation. In this regard, our study suggests that strict intervention should be implemented by the Government in the subsequent period of this outbreak. We believe that the findings obtained from this study can provide fruitful insights in framing policies regarding the control of COVID-19 in India.

\section*{Acknowledgements}

Abhishek Senapati is supported by the research fellowship from Council of Scientific \& Industrial Research, India (Grant no: 09/093(0167)/2015/EMR-I), Government of India.

\section*{Conflict of interest}

The authors declare that there is no conflict of interest.

%\clearpage

%\bibliographystyle{plain}
\bibliographystyle{elsarticle-num}
\bibliography{Corona_India}

\end{document}